\pdfoutput=1
\documentclass[aps,prb,reprint,superscriptaddress,longbibliography,nofootinbib]{revtex4-1}

\usepackage{graphicx}
\usepackage{amsmath}
\usepackage{amssymb}
\usepackage{amsfonts}
\usepackage{dcolumn}
\usepackage{dsfont}
\usepackage{latexsym}
\usepackage{rotating}
\usepackage{color}
\usepackage{latexsym}
\usepackage{bbm}
\usepackage{subfigure}
\usepackage{float}
\usepackage{epsfig}
\usepackage{psfrag}
\usepackage{natbib, hyperref}
\usepackage{bm}
\usepackage{amsthm}
\usepackage{eucal}
\usepackage{mathrsfs}
\usepackage{url}
\usepackage{braket}
\usepackage{ulem}
\usepackage{calligra}
\usepackage[T1]{fontenc}
\usepackage[utf8]{inputenc}
\usepackage{newunicodechar}
\usepackage{marvosym}
\usepackage[absolute]{textpos}
\usepackage[cal=cm]{mathalfa}
\usepackage{soul}

\usepackage{color}

\usepackage{hyperref}
\hypersetup{
colorlinks=true,final=true,
        linkcolor=blue,
        citecolor=blue,
        filecolor=blue,
        urlcolor=blue,
}

\begin{document}
\setlength{\abovedisplayskip}{3pt}
\setlength{\belowdisplayskip}{3pt}

\title{Planar topological Hall effect from conical spin spirals}

\author{Narayan Mohanta}
\email{Email: mohantan@ornl.gov}
\affiliation{Material Science and Technology Division, Oak Ridge National Laboratory, Oak Ridge, TN 37831, USA}
\author{Satoshi Okamoto}
\affiliation{Material Science and Technology Division, Oak Ridge National Laboratory, Oak Ridge, TN 37831, USA}
\author{Elbio Dagotto}
\affiliation{Material Science and Technology Division, Oak Ridge National Laboratory, Oak Ridge, TN 37831, USA}
\affiliation{Department of Physics and Astronomy, The University of Tennessee, Knoxville, TN 37996, USA}

\begin{abstract}
Planar Topological Hall effect (PTHE) is known to arise from in-plane  skyrmion tubes with an external magnetic field applied in the plane of the charge current. Here, we propose that the PTHE can robustly appear also from an unexpected source, the conical spin spiral, realizable in a variety of materials. We show that for both in-plane skyrmion tubes and conical spin spirals, the PTHE is two-fold symmetric with respect to the magnetic-field angle, while it is four-fold symmetric for 90$^{\circ}$-rotated domains of conical spin spirals. We predict that the symmetry and magnetic-field response of the PTHE can distinguish between the two scenarios, unambiguously probing the nature of the antisymmetric spin-exchange interaction and the resulting magnetic texture.
\end{abstract}

\maketitle

\section{Introduction}
\vspace{-3mm}
Transport of electrons in chiral magnetic textures has emerged as a central topic towards exciting new physics and spintronics applications~\cite{Nagaosa_NNat2013,Barron_NMat2008,Fert_NRevMat2017}. In noncentrosymmetric magnets and oxide interfaces, the chiral magnetic textures typically originate from Dzyaloshinskii-Moriya (DM) -type antisymmetric exchange interactions~\cite{Rossler_Nature2006,Balents_PRL2014}. A non-collinear spin texture, such as the skyrmion crystal (SkX), possesses a finite scalar spin chirality which produces an emergent magnetic field and influences the electronic transport, inducing topological Hall effect in metals~\cite{Neubauer_PRL2009,Muhlbauer_Science2009,Yu_Nature2010,Heinze_NPhys2011,Yu_NMat2011,Kanazawa_PRL2011,Huang_PRL2012,Schulz_Nphys2012,Nagaosa_NNat2013,Park_NNat2014,Matsuno_SciAdv2016,Nakamura_JPSJ2018,Vistoli_NPhys2019}. \\
\indent In the planar Hall geometry, with coplanar magnetic field and charge current, a  field-angular modulation of the planar topological Hall conductivity was proposed to originate from in-plane skyrmion tubes (IST) when the DM interaction has a finite component perpendicular to the plane where the magnetic field is applied, as in noncentrosymmetric magnets~\cite{Tokura_JPSJ2015}. In several compounds and heterostructures with broken structural inversion symmetry, the DM interaction is intrinsically two-dimensional (2D) in nature~\cite{Banerjee_NPhys2013,Matsuno_SciAdv2016,Li_ACS2019,Jiang_PRB2019} and lies in the plane of the applied magnetic field. In this scenario, PTHE is unanticipated since the formation of IST is not possible.

\indent In this paper, we show that conical spin spirals (CSS), the underlying magnetic texture with DM vectors parallel to the plane where the magnetic field is applied, can also generate a significant PTHE. This finding opens opportunities to elucidate transport properties originating in chiral magnetic textures beyond skyrmions. We find that the PTHE from the CSS is two-fold symmetric with respect to the magnetic-field angle -- similar to that from the IST -- when only one of the two orthogonal stripe arrangements is present. However, in real magnetic materials, both stripe arrangements often coexist and form labyrinth-like metastable domains, induced by defect pinning. The PTHE in the case of a mosaic of 90$^{\circ}$-rotated stripe domains is four-fold symmetric, clearly distinguishable from the PTHE arising from the IST. Based on the symmetry and magnetic-field dependence of the PTHE, we formulate strategies to characterize the nature of the antisymmetric spin-exchange interaction which is often difficult to understand only from indirect experimental evidences. Moreover, the CSS discussed here has the following potential technological advantages: (i) the PTHE at low field regimes is proportional to the applied magnetic field and, thus, could effectively amplify the effect of magnetic fields to the Hall conductivity significantly. (ii) the CSS has higher controllability than IST in terms of the spin helicity. For the CSS, one could select one of the two spin helices by applying a magnetic field during the cooling cycle, while for the IST the spin helicity is fixed by the underlying crystal structure.
\begin{figure}[t]
\begin{center}
\epsfig{file=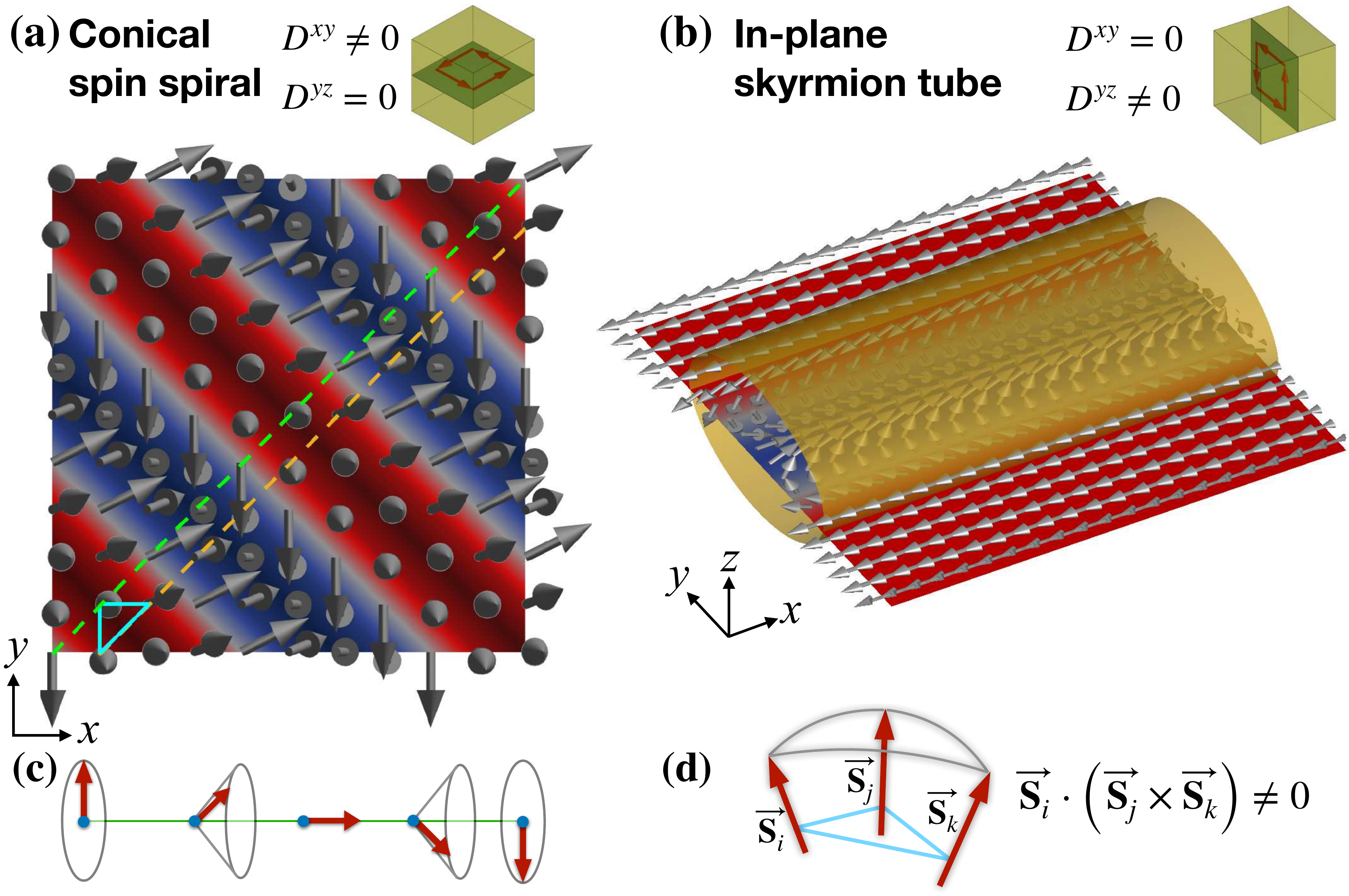,trim=0.0in 0.1in 0.0in 0.0in,clip=true, width=85mm}
\caption{Schematic spin arrangements in a 2D plane for (a) the CSS originating from in-plane DM vectors and (b) the IST originating from out-of-plane DM vectors. (c) Conical spin arrangement along the green (or yellow) dashed line in (a). (d) In both plots (a) and (b), three spins on some triangular plaquettes, as shown by the cyan triangle in (a), subtend a finite solid angle in real space, creating a finite scalar spin chirality, the source of the PTHE.}
\label{fig1}
\vspace{-4mm}
\end{center}
\end{figure}

\begin{figure*}[t]
\begin{center}
\epsfig{file=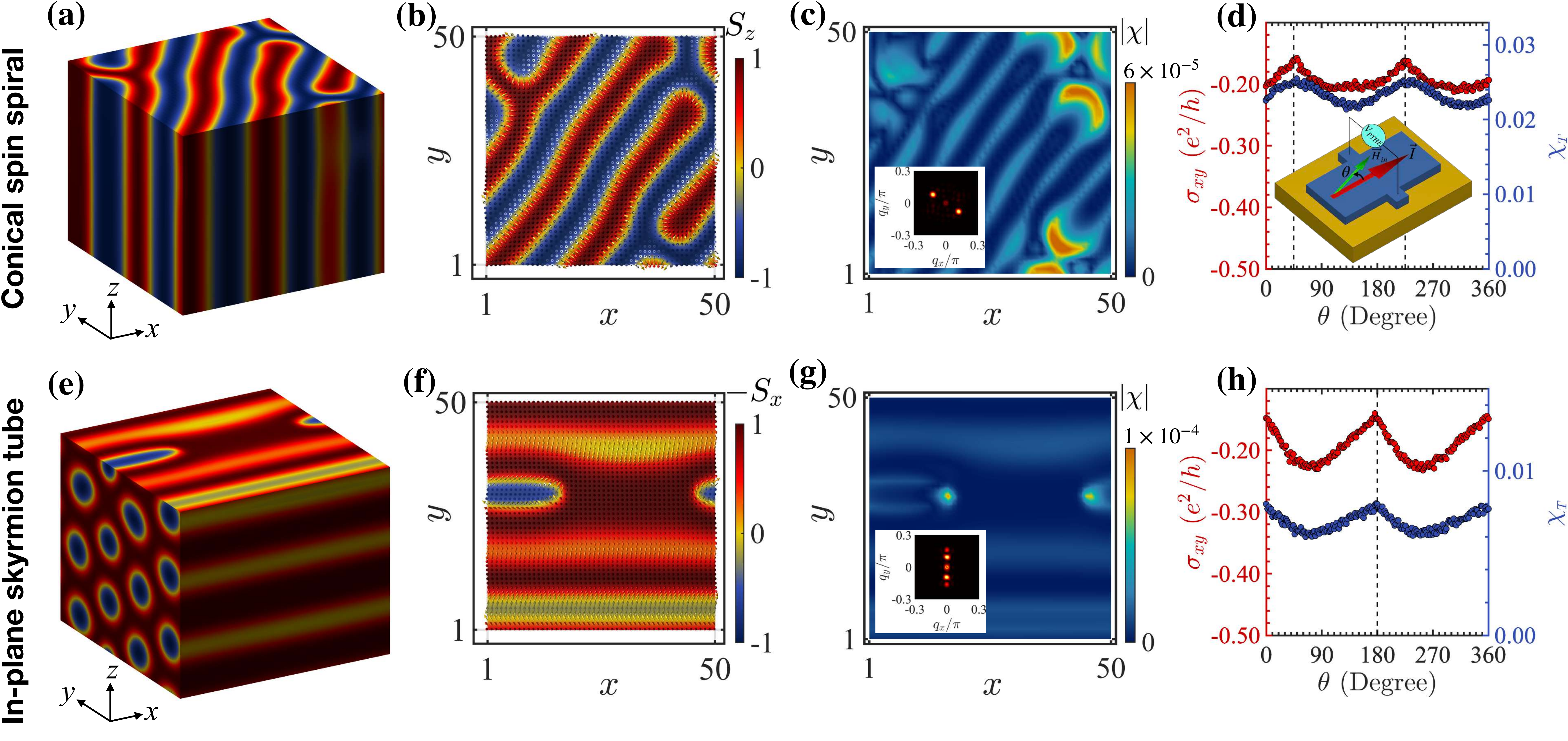,trim=0.3in 0.3in 0.0in 0.0in,clip=false, width=170mm}
\caption{Spin texture and planar Hall response of the CSS (a - d) and IST (e - h). Colormap of (a) $S_z$ profile, (e) $-S_x$ profile in a 50$\times$50$\times$50 cubic system with the DM vectors lying in the (a) $x$-$y$ plane and (e) $y$-$z$ plane. A magnetic field of strength $H_{in} \!=\! 0.15J$ is applied along the $x$ direction in both cases. (b),(f) Spin texture on the $z \!=\! 50$ layer of the cubic system, revealing the CSS and the IST textures. (c),(g) Profile of the scalar spin chirality $\chi$ for the spin textures in (b) and (f), respectively. Insets in (c) and (g) show the Bragg intensity profile $I(\mathbf{q})$ in the 2D Brillouin zone. (d),(h) Variation of the MC-averaged transverse Hall conductivity $\sigma_{xy}$ (left $y$ axis) and total scalar spin chirality $\chi_{_T}$ (right $y$ axis) with the field angle $\theta$, for (d) the CSS and (h) the IST textures. Inset in (d) shows the geometry of the PTHE with a charge current assumed to be running along the $x$ direction, the Hall voltage measured along the transverse direction, and the magnetic-field angle $\theta$ determined with respect to the current direction. Parameters used are $J \!=\! 1$, $S \!=\! 1$, $D \!=\! 0.5$, and $T \!=\! 0.001$.}
\label{fig2}
\vspace{-4mm}
\end{center}
\end{figure*}
In Fig.~\ref{fig1}, we show schematically the two spin textures under consideration, \textit{viz.} the CSS (with DM vectors lying in the $x$-$y$ plane, \textit{i.e.} $\mathbf{D}^{xy}$$\neq$$0$, $\mathbf{D}^{yz}$=$0$) and the IST (with DM vectors lying in the $y$-$z$ plane, \textit{i.e.} $\mathbf{D}^{xy}$=$0$, $\mathbf{D}^{yz}$$\neq$$0$). The key difference is that the DM vectors lie in different planes. In both cases, a finite scalar spin chirality $\chi_{ijk}$=$\mathbf{S}_i \cdot (\mathbf{S}_j \times \mathbf{S}_k)$, the origin of the PTHE, emerges in some regions of the 2D plane. A finite solid angle subtended by $\hat{n}(\mathbf{r})$=$\mathbf{S(\mathbf{r})}/|\mathbf{S(\mathbf{r})}|$, the orientation of the localized spin $\mathbf{S(\mathbf{r})}$, over an infinitesimal loop in space in the continuum limit generates an emergent magnetic field, which can be expressed in two dimensions as $B_{z}^{e}$=$\frac{\hbar}{2e} \hat{n}\cdot \Big( \frac{\partial \hat{n} }{\partial x} \times  \frac{\partial \hat{n} }{\partial y} \Big)$~\cite{Schulz_Nphys2012,Nagaosa_PhilTrans2012,Nagaosa_NNat2013}. In our context of the CSS or the IST, the scalar spin chirality $\chi$ and the emergent magnetic field $B_{z}^{e}$ are also finite, and a PTHE is expected to appear in an in-plane magnetic field.

\section{Model and formalism}
\vspace{-3mm}
The transverse Hall response in the planar Hall geometry [Fig.~\ref{fig2}(d) inset] is studied here using the chiral magnetic texture of the top layer of a three-dimensional (3D) cubic system having isotropic ferromagnetic exchange interactions and 2D DM vectors. At zero field and a low temperature, the spin texture is obtained by using Metropolis Monte Carlo (MC) annealing~\cite{NM_PRB2019,SM}, formulated using the spin Hamiltonian
\begin{align}
{\cal{H}}=&-J\sum_{\langle  ij \rangle} \mathbf{S}_i \cdot \mathbf{S}_j
- \sum_{i} \mathbf{H}_{in} \cdot \mathbf{S}_i  
-  \sum_{\langle  ij \rangle} \mathbf{D}_{ij}^{\alpha \beta} \cdot (\mathbf{S}_{i} \times \mathbf{S}_{j} ) \nonumber,
\end{align}
where $J>0$ is the strength of the ferromagnetic Heisenberg exchange, $i$ and $j$ indicate nearest-neighbor lattice sites in 3D space, $\mathbf{H}_{in}$ is the strength of the applied in-plane magnetic field, and $\mathbf{D}_{ij}^{\alpha \beta}$=$D(\hat{\alpha} \times \hat{\beta})\times \hat{r}_{ij}$ represents the DM vector between $i$ and $j$ lying in the $\alpha-\beta$ plane, with $D$ being the strength of the DM interaction and $\hat{r}_{ij}$ the unit vector between $i$ and $j$. Below we consider two cases separately: (a) $|\mathbf{D}_{ij}^{xy}| \neq 0$, $|\mathbf{D}_{ij}^{yz}| \!=\! 0$ and (b) $|\mathbf{D}_{ij}^{xy}| \!=\! 0$, $|\mathbf{D}_{ij}^{yz}| \neq 0$. When adding the in-plane field, the spin textures were obtained by performing the MC evolution starting from the previously MC converged zero-field low-temperature spin configuration, without changing the temperature. See the Supplemental Material~[\onlinecite{SM}] for details. Each field angle $\theta$ was measured with respect to the $x$ direction.\\
\indent With the DM vectors lying in the $x$-$y$ plane, Fig.~\ref{fig2}(a) shows the CSS texture, spontaneously generated by the MC method in a $50$$\times$$50$$\times50$ lattice with periodic boundary conditions at $H_{in} \!=\! 0.15J$ and $\theta \!=\! 0^{\circ}$. The $z\!=\!50$ layer, shown in Fig.~\ref{fig2}(b), exhibits a nonzero scalar spin chirality profile, presented in Fig.~\ref{fig2}(c), calculated via $\chi(\mathbf{r}_{i}) \!=\! \sum_{ijk \in p} | \mathbf{S}_i \cdot (\mathbf{S}_j \times \mathbf{S}_k) |$, where $i$, $j$ and $k$ are the three sites of a triangular plaquette $p$ in the underlying square lattice. The Bragg intensity profile $I(\mathbf{q}) \!=\! (1/N)\sum_{ij} \langle \mathbf{S}_{i} \cdot \mathbf{S}_{j} \rangle e^{-i\mathbf{q} \cdot (\mathbf{r}_{i}-\mathbf{r}_{j}) }$, where $N$ is the total number of lattice sites on the $z\!=\!50$ layer, shows a small peak at $\mathbf{q} \!=\! \mathbf{0}$ indicating a small spin polarization [Fig.~\ref{fig2}(c) inset], a characteristic feature distinguishing a CSS from the regular spin spiral which appears at $H_{in} \!=\! 0$.\\
\indent To calculate the transverse Hall conductivity at the top layer of the 3D magnetic system, we consider the double-exchange Hamiltonian
\begin{align}
{\cal{H}}_{DE}=-t\sum_{\langle ij \rangle} c_{i \sigma}^{\dagger}c_{j \sigma} -J^{\prime} \sum_{i,\sigma,\sigma^{\prime}} (\mathbf{S}_i \cdot \boldsymbol{\sigma}_{\sigma \sigma^{\prime}} ) c_{i \sigma}^{\dagger}c_{i \sigma^{\prime}} \nonumber, 
\label{HDE}
\end{align}
where $t$ is the nearest-neighbor electron hopping energy and $J^{\prime}$ is the exchange coupling strength between the itinerant electron $\boldsymbol{\sigma}$ and localized spins $\mathbf{S}_i$. The localized spin configuration is generated using MC simulations. The Hall conductivity is obtained via the Kubo formula
\begin{align}
\sigma_{xy}\!=\!\frac{e^2}{h}\frac{2\pi}{N} \! \sum_{\epsilon_m \neq \epsilon_n} \! \frac{f_m-f_n}{(\epsilon_m-\epsilon_n)^2+\eta^2} \text{Im}\Big( \langle m | \hat{j_x} | n \rangle \langle n | \hat{j_y} | m \rangle \Big) \nonumber,
\end{align}
where $f_m$ is the Fermi function at temperature $T$ and energy $\epsilon_m$, $\hat{j}_x$ ($\hat{j}_y$) is the current operator along the $x$ ($y$) direction, $| m \rangle$ is the $m^{\rm th}$ eigenstate of ${\cal{H}}_{DE}$, and $\eta$ is the relaxation rate. We used $t \!=\! 1$, $J^{\prime} \!=\! 1$, and $\eta \!=\! 0.1$ for the results presented here, with no qualitative difference in the description for other choices, as elaborated before~\cite{NM_PRB2019}. MC averaging of $\sigma_{xy}$ was performed by considering 50 different spin textures realizations, obtained at a low temperature $T$=$0.001J$.

\section{PTHE from conical spin spirals}
\vspace{-3mm}
The Hall conductivity $\sigma_{xy}$ for the CSS texture, versus the magnetic-field angle $\theta$, is in Fig.~\ref{fig2}(d), displaying an anisotropic behavior. The anisotropy arises from the stripes being oriented along one of the diagonals. The field angle where $|\sigma_{xy}|$ is maximized (minimized) corresponds to a field perpendicular (parallel) to the CSS stripes in Fig.~\ref{fig2}(b). The total scalar spin chirality $\chi_{_T}$ [see right $y$ axis in Fig.~\ref{fig2}(d)] also displays a similar anisotropic behavior, confirming the unexpected appearance of a finite PTHE from the CSS texture. Since the DM vectors lie entirely in the $x$-$y$ plane, the $\theta$-dependence of $\sigma_{xy}$ is  identical for different thicknesses of the material. In other words, for 2D DM vectors in the plane of the magnetic field, the PTHE is independent of the thickness of the magnetic material. The PTHE, at fixed magnetic field, is found to decrease with increasing the periodicity of the CSS texture, that is primarily governed by the DM interaction amplitude, as described in the Supplemental Material~[\onlinecite{SM}]. The two-fold symmetry of the PTHE  exists as long as the CSS texture is homogeneous {\textit i.e.} without rotated domains. This phenomenon should appear at perovskite-oxide interfaces (\textit{e.g.} manganite-iridate interfaces) where the interface DM interaction originates from Rashba spin-orbit coupling~\cite{Banerjee_NPhys2013,Balents_PRL2014,NM_PRB2019,Mohanta_arXiv2020Spinwave}. 

\section{PTHE from in-plane skyrmion tubes}
\vspace{-3mm}
\begin{figure}[t]
\begin{center}
\epsfig{file=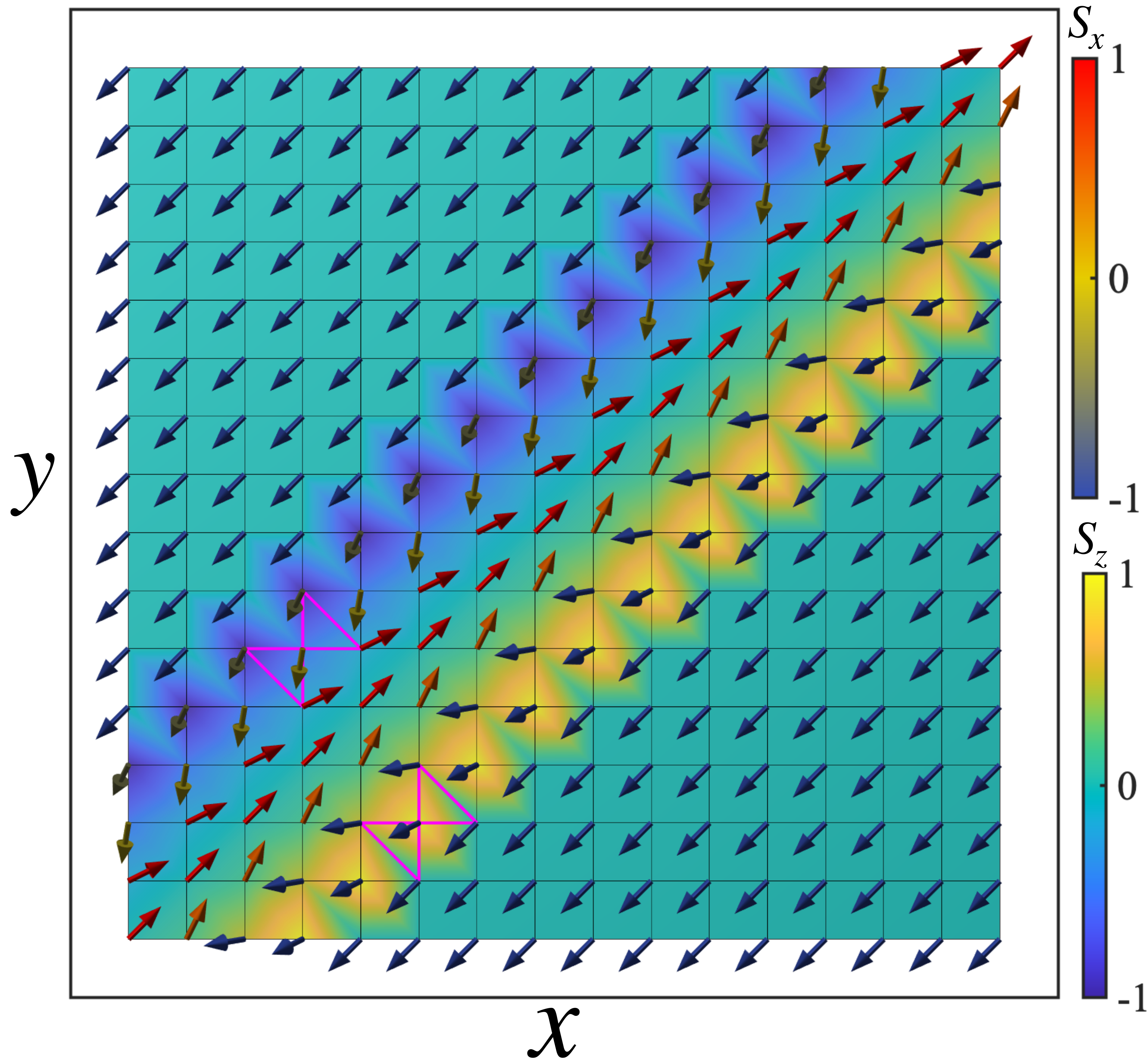,trim=0.38in 0.1in 0.0in 0.0in,clip=false, width=50mm}
\caption{Schematic representation of an in-plane tube of Bloch-type skyrmion, propagating along a diagonal of the 2D plane. The scalar spin chirality on the purple triangular plaquettes is finite and it can give rise to a PTHE at $T\!=\!0$.}
\label{fig3}
\vspace{-4mm}
\end{center}
\end{figure}
\begin{figure}[b]
\begin{center}
\epsfig{file=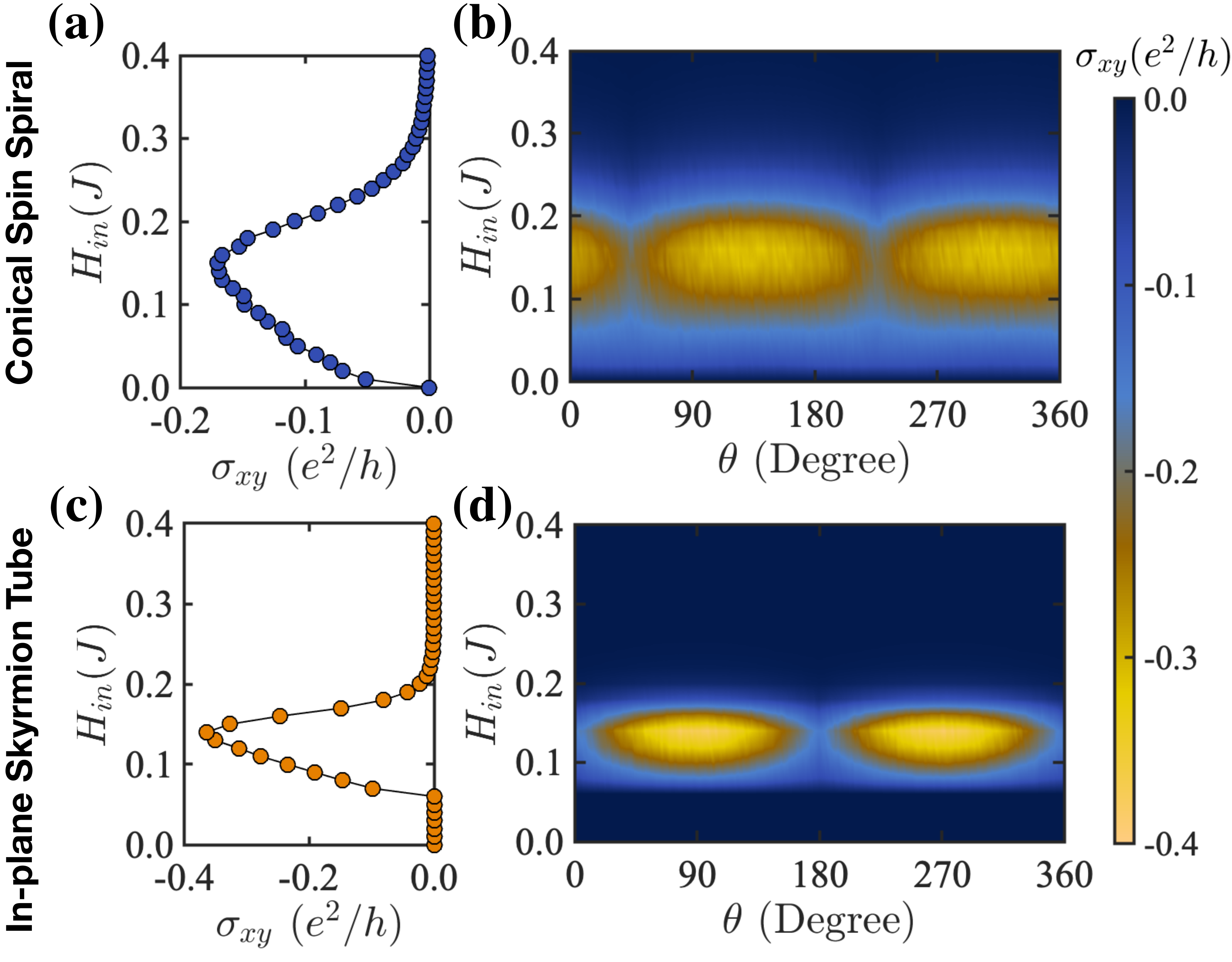,trim=0.38in 0.1in 0.0in 0.0in,clip=false, width=76mm}
\caption{MC-averaged transverse Hall conductivity $\sigma_{xy}$ vs magnetic field amplitude $H_{in}$ for (a) a CSS texture at a field angle $\theta \!=\! 0^{\circ}$  and (c) an IST texture at $\theta \!=\! 180^{\circ}$, on the $z$=$50$ layer of the 50$\times$50$\times$50 lattice. Plots (b) and (d) show the variation of $\sigma_{xy}$ for (b) a CSS texture and (d) an IST texture with field angle $\theta$ and field amplitude $H_{in}$. Other parameters are as in Fig.~2.}
\label{fig4}
\vspace{-4mm}
\end{center}
\end{figure}
Now consider a second scenario, discussed earlier in Ref.~[\onlinecite{Tokura_JPSJ2015}]. In noncentrosymmetric magnets~\cite{Ezawa_PRB2011,Buhrandt_PRB2013,Wilson_PRB2014} and layered ferromagnets~\cite{Park_arXiv2019,You_PRB2019}, the DM interaction might have a finite component out of plane and an anisotropy, sizable in thin-films, that stabilizes the SkX phase in a plane perpendicular to the magnetic field. The skyrmions form tubes with the tube axis perpendicular to the plane of the DM vectors.  Therefore, some of these skyrmion tubes -- those that coincide with the top layer of the 3D lattice where the magnetic field is applied -- should produce a PTHE, as conjectured in the context of MnSi~\cite{Tokura_JPSJ2015}. Consider the DM vectors lying in the  $y$-$z$ plane. For a magnetic field applied along the $x$ axis, a N\'{e}el-type SkX is stabilized in the $y$-$z$ plane, as in Fig.~\ref{fig2}(e). The IST on the top layer ($z \!=\! 50$) is shown in Fig.~\ref{fig2}(f). The scalar spin chirality $\chi$, see Fig.~\ref{fig2}(g), is finite near the skyrmion tubes, oriented along the $x$ direction. A weak canting in the N\'{e}el-type IST induced by a finite $T$ is crucial for a nonzero PTHE in the considered geometry for the IST. However, a Bloch-type IST (found in MnSi), oriented diagonally, as in Fig.~\ref{fig3}, is expected to produce a robust PTHE at $T\!=\!0$. Note that the nonzero chirality produced by CSS survives at $T\!=\!0$. Fig.~\ref{fig2}(f) and (g) also show significant contributions from the magnetic monopoles at the ends of the chiral bobbers (see \textit{e.g.} Ref.~\onlinecite{Zheng_NNat2018,Ahmed_PRM2018}) that happen to exist on the top layer. The Bragg intensity profile $I(\mathbf{q})$ for the IST texture is in Fig.~\ref{fig2}(g) (inset). The transverse Hall conductivity $\sigma_{xy}$, plotted in Fig.~\ref{fig2}(h) with respect to $\theta$, presents an enhancement when the magnetic field is applied at an angle away from the IST orientation direction \textit{i.e.} along the $x$ axis.

\section{Magnetic-field variation of PTHE}
\vspace{-3mm}
To explore the differences in the PTHE between the CSS and IST cases, we now study the variation of PTHE with the external magnetic-field amplitude. The SkX phase, with $|\mathbf{D}_{ij}^{yz}| \! \neq  \!0$, $|\mathbf{D}_{ij}^{xy}| \!=\! 0$, is stabilized in our cubic magnetic system within a range of field amplitudes, when applied parallel to the $x$ axis. Consequently, in topological Hall measurements, the formation of the SkX phase is indicated by an enhancement in the topological Hall conductivity within a sharp range of magnetic fields. In Fig.~\ref{fig4}, we compare the field-dependence of the PTHE response from the two kinds of chiral spin textures. \\
\begin{figure}[t]
\begin{center}
\epsfig{file=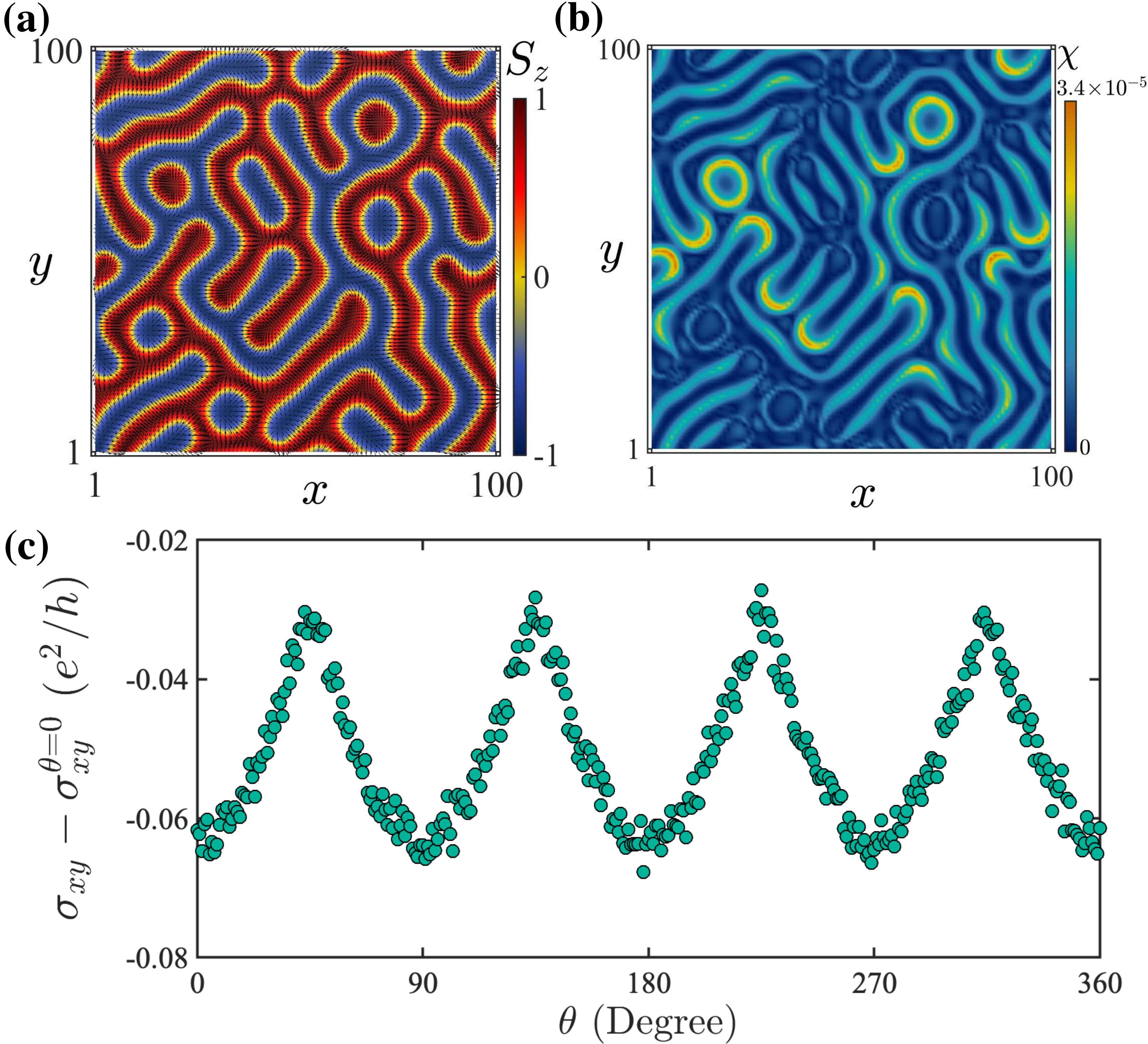,trim=0.0in 0.1in 0.0in 0.0in,clip=false, width=85mm}
\caption{(a) MC spin texture of a CSS texture on the top layer of a 100$\times$100$\times$20 cubic lattice, showing coexisting domains of two orthogonal CSS arrangements in an in-plane magnetic field of strength $H_{in} \!=\! 0.06J$, applied along the $x$ direction. (b) Profile of the scalar spin chirality $\chi$ for the CSS domains in panel (a). (c) MC-averaged transverse Hall conductivity $\sigma_{xy}$, after subtracting the zero-field value, plotted with respect to the angle $\theta$ of the in-plane magnetic field, showing a four-fold symmetry of the PTHE. Other parameters are as in Fig.~2.}
\label{fig5}
\vspace{-4mm}
\end{center}
\end{figure}
\indent For the CSS, the transverse Hall conductivity $\sigma_{xy}$ increases with the magnetic field in the low-field regime ($H_{in}$$\lesssim$$0.16J$), and decreases in the high-field regime ($H_{in}$$\gtrsim$$0.16J$), see Fig.~\ref{fig4}(a). Such a non-monotonic field-dependence appears because the CSS phase is stabilized within a broader range of fields, in between the normal spin spiral phase at zero field and the in-plane ferromagnetic phase at very large fields, with both of these limits having zero topological Hall conductivity. Figure~\ref{fig4}(b) shows a similar enhancement in $\sigma_{xy}$ at all field angles $\theta$, with minima near $\theta \!=\! 45^{\circ}$ and $\theta \!=\! 225^{\circ}$ \textit{i.e.} the alignment direction of the CSS stripes. \\
\indent Conversely, $\sigma_{xy}$ for the IST enhances rapidly and remains nonzero only within a sharply-defined field range $0.08J$$\lesssim$$H_{in}$$\lesssim$$0.2J$. The field range for the IST texture is smaller than for the CSS texture. Figure~\ref{fig4}(d) shows the $\theta$-dependence of the field-variation in $\sigma_{xy}$, two-fold symmetric, with the minima located at $\theta \!=\! 0^{\circ}$ and $\theta \!=\! 180^{\circ}$, fixed by the orientation  of the ISTs.

\section{Four-fold symmetric PTHE from 90$^{\circ}$-rotated domains}
\vspace{-3mm}
In general, the spin spirals have two orthogonal solutions which often coexist to form maze-like domain structures, also known as labyrinth domains. These metastable domains can be induced by pinning of the spin spirals by impurities. The existence of such 90$^{\circ}$-rotated domains was reported in several compounds (both bulk and heterostructures), including thin ferromagnetic films~\cite{Wolfe_PRA1992,Kashuba_PRL1993,Portmann_Nature2003,Saratz_PRL2010} and noncentrosymmetric magnets~\cite{Muhlbauer_Science2009,Woo_NMater2016}, in the absence of any magnetic field. In Fig.~\ref{fig5}(a), we show the spin texture with  domains of CSS textures formed on the top layer in a 100$\times$100$\times$20 lattice -- with a larger lateral dimension than the previously-discussed lattice -- at low temperature $T$=$0.001J$ and in-plane magnetic field $H_{in}$=$0.06J$, with the DM vectors lying in the $x$-$y$ plane. These domains were obtained as metastable solutions of the MC evolutions. Typical spatial profiles of the scalar spin chirality $\chi$, as in Fig.~\ref{fig5}(b), present pronounced features at the boundaries between the 90$^{\circ}$-rotated domains. The finite scalar spin chirality at these boundaries can produce a topological Hall effect even at zero magnetic field, as described in the Supplemental Material~[\onlinecite{SM}] (see, also, references ~[\onlinecite{Kato_PRApp2019}] and [\onlinecite{Nakane_PRB2020}] therein).\\  
\indent In an in-plane magnetic field, after subtracting the zero-field contribution arising from the domain boundaries, we find that the transverse Hall conductivity, MC-averaged over several domain configurations, exhibits a four-fold symmetry with respect to the field angle $\theta$, as in Fig.~\ref{fig5}(c). The four minima in $|\sigma_{xy}$--$\sigma_{xy}$$(H_{in} \!=\! 0)|$ appears at $\theta \!= \!(2n-1)\times45^{\circ}$, $(n\! =\! 1,2,3,4)$. For the CSS texture having only one diagonal arrangement, as in Fig.~\ref{fig2}(a), if the applied magnetic field induces a reorientation of the stripes~\cite{Fin_PRB2015,Kent_JPCM2001}, $\sigma_{xy}$ is also expected to reveal a four-fold symmetric $\theta$-dependence. Such rotated domains does not appear in the IST case since it appears perpendicular to the plane of the DM vectors, typically fixed in a given material. Therefore, the observation of such a four-fold symmetric PTHE can convincingly identify the CSS texture and exclude the IST texture. This approach to identify magnetic textures by looking at the PTHE signal is best suited for chiral magnetic textures which are periodic and homogeneous since the four-fold symmetric PTHE appears from the coexistence of the 90$^{\circ}$-rotated CSS domains.

\section{Discussion}
\vspace{-3mm}
We discussed 2D DM interactions and their resulting spiral and skyrmions N\'{e}el-type textures. This could be realized in effective 2D lattices such as  oxide interfaces or bulk 3D systems with DM vectors lying identically in parallel 2D planes. Moreover, for a DM interaction with both in-plane and out-of-plane components, the resulting textures are Bloch type~\cite{Neubauer_PRL2009,Muhlbauer_Science2009,Yu_Nature2010}. The  present description of PTHE is also applicable for the Bloch-type magnetic textures.\\
\indent To conclude, we propose the emergence of topological Hall effect from CSS textures, when the magnetic field, charge current, and DM vectors all lie in the same plane. The PTHE generically develops from the real-space Berry curvature of the CSS texture and it appears in several magnetic materials. We outline the distinguishing features of the planar topological Hall response from two possible sources, namely the in-plane skyrmions and the conical spin spirals, which are important to unambiguously probe chiral magnetic textures and identify the nature of the antisymmetric spin-exchange interaction.

\section*{Acknowledgements}
\vspace{-3mm}
This work was supported by the U.S. Department of Energy (DOE), Office of Science, Basic Energy Sciences (BES), Materials Sciences and Engineering Division.

\vspace{-3mm}

%

\end{document}